\documentclass[runningheads,a4paper]{llncs}

\usepackage{amssymb}
\usepackage{amsmath}
\usepackage{graphicx}

\usepackage{url}
\urldef{\mailsa}\path|niklas.baumstark@gmail.com|
\urldef{\mailsb}\path|{guyb,jshun}@cs.cmu.edu|

\usepackage{listings}
\usepackage{hyperref}
\usepackage[framemethod=tikz]{mdframed}
\usepackage[group-separator={,}]{siunitx}
\usepackage{multirow}

\newcommand\mylstcaption{}

\usepackage[space]{cite}

\begin{document}

\mainmatter  

\def\doctitle{Efficient Implementation of a Synchronous Parallel Push-Relabel Algorithm}
\title{\doctitle
\thanks{This is a longer version of the paper appearing in the proceedings of the \emph{European Symposium on Algorithms}, 2015. The final publication is available at \url{link.springer.com}}
}
\titlerunning{\doctitle}

\author{Niklas Baumstark$^1$, Guy Blelloch$^2$, Julian Shun$^2$}
\authorrunning{\doctitle}

\institute{$^1$Institute of Theoretical Informatics, Karlsruhe Institute of Technology, Germany\\
\mailsa \\
$^2$Computer Science Department, Carnegie Mellon University, USA\\
\mailsb
}

\toctitle{\doctitle}
\tocauthor{Niklas Baumstark, Guy Blelloch, and Julian Shun}
\maketitle

\begin{abstract}
Motivated by the observation that FIFO-based push-relabel algorithms are able to
outperform highest label-based variants on modern, large maximum flow problem
instances, we introduce an efficient implementation of the algorithm that uses
coarse-grained parallelism to avoid the problems of existing parallel
approaches. We demonstrate good relative and absolute speedups of our algorithm
on a set of large graph instances taken from real-world applications.
On a modern 40-core machine, our parallel implementation outperforms existing
sequential implementations by up to a factor of 12 and other parallel
implementations by factors of up to 3.
\end{abstract}

\section{Introduction}

The problem of computing the maximum flow in a network plays an important role
in many areas of research such as resource scheduling, global optimization and
computer vision. It also arises as a subproblem of other optimization tasks
like graph partitioning. There exist near-linear approximate algorithms for the
problem~\cite{Kelner2013}, but exact solutions can in practice be found
even for very large instances using modern algorithms.  It is only natural to
ask how we can exploit readily available multi-processor systems to further
reduce the computation time.
While a large fraction of the prior work has focused on distributed and
parallel implementations of the algorithms commonly used in computer vision,
fewer publications are dedicated to finding parallel algorithms that solve the
problem for other graph families.

To assess the practicality of existing algorithms, we collected a number of
benchmark instances. Some of them are taken from a common benchmark suite for
maximum flow and others we selected specifically to represent various
applications of maximum flow.
Our experiments suggest that Goldberg's \emph{hi\_pr} program (a highest
label-based push-relabel implementation) which is often used for comparison in
previous publications is not optimal for most of the graphs that we studied.
Instead, push-relabel algorithms processing active vertices in first-in-first-out
(FIFO) order seems to be better suited to these graphs, and at the same time happen
to be amenable for parallelization.
We proceeded to design and implement our own shared memory-based parallel
algorithm for the maximum flow problem, inspired by an old algorithm
and optimized for modern shared-memory platforms.  In contrast to previous parallel
implementations we try to keep the usage of atomic CPU instructions to a minimum.
We achieve this by employing coarse-grained synchronization to rebalance the
work and by using a parallel version of global relabeling instead of running it
concurrently with the rest of the algorithm.

We are able to demonstrate good speedups on the graphs in our benchmark suite,
both compared to the best sequential competitors, where we achieve speedups of
up to 12 with 40 threads, and to the most recent parallel solver, which we often
outperform by a factor of three or more with 40 threads.

\section{Preliminaries and Related Work}

We consider a directed graph $G$ with vertices $V$, together with a designated
source $s$ and sink $t$, where $s \neq t \in V$ as well as a capacity function
$c : V \times V \rightarrow \mathbb{R}_{\geq 0}$.  The set of edges is $E = \{
(v, w) \in V \times V \ | \ c(v, w) > 0 \}$. We define $n = |V|$ and $m = |E|$.
A \emph{flow} in the graph is a function $f: E \rightarrow \mathbb{R}$ that is
bounded from above by the capacity function and respects the \emph{flow conservation} and
\emph{asymmetry} constraints

\begin{align}
\forall w \in V: && \sum_{(v,w) \in E, v \neq w} f(v, w) &= \sum_{(w,x) \in E, w \ne x} f(w, x)&& \\
\forall v, w \in V: &&  f(v, w) &= -f(w, v)&&
\end{align}

We define the \emph{residual graph} $G_f$ with regard to a specific flow $f$ using the
residual weight function $c_f(v, w) = c(v, w) - f(v, w)$. The set of residual
edges is just $E_f = \{ (v, w) \in V \times V \ |\ c_f(v, w) > 0 \}$. The \emph{reverse
residual graph} $G^R_f$ is the same graph with each edge inverted.

A \emph{maximum flow} in $G$ is a flow that maximizes the \emph{flow value}, i.e. the
sum of flow on edges out of the source. It so happens that a flow is maximum if and only
if there is no path from $s$ to $t$ in the residual graph $G_f$
\cite[Corallary 5.2]{Ford1962}.
The maximum flow problem is the problem of finding such a flow function. It
is closely related to the minimum cut problem, which asks for a disjoint
partition $(S \subset V, T \subset V)$ of the graph with $s \in S, t \in T$
that minimizes the cumulative capacity of edges that cross from $S$ to $T$.  It
can be shown that the value of a maximum flow is equal to the value of
a minimum cut and a minimum cut can be easily computed from a given maximum
flow in linear time as the set of vertices reachable from the source in the
residual graph \cite[\S5]{Ford1962}.

\subsection{Sequential Max-Flow and Min-Cut Computations}

Existing work related to the maximum flow problem is generally split into two
categories: work on algorithms specific to computer vision applications and
work on general-purpose algorithms. Most of the algorithms that work well for the
type of grid graphs found in computer vision tend to be inferior for other graph
families and vice versa~\cite[Concluding Remarks]{Goldberg2011}. In this paper we
aim to design a general-purpose algorithm that performs reasonably well on all
sorts of graphs.

Traditional algorithms for the maximum flow problem typically fall into one
of two categories: \emph{Augmenting path-based} algorithms directly apply the
\emph{Ford--Fulkerson theorem}~\cite[Corallary 5.2]{Ford1962} by incrementally
finding augmenting paths from $s$ to $t$ in the residual graph and increasing
the flow along them. They mainly differ in their methods of finding augmenting
paths. Modern algorithms for minimum cuts in computer vision applications such
as \cite{Goldberg2011} belong to this family.
\emph{Preflow-based} algorithms do not maintain a valid flow during their
execution and instead allow for vertices to have more incoming than outgoing
flow. The difference in flow on in-edges and out-edges of a vertex is called
\emph{excess}. Vertices with positive excess are called \emph{active}.
A prominent member of this family is the classical \emph{push-relabel} algorithm
due to Goldberg and Tarjan~\cite{Goldberg1988}. It maintains vertex labels that
estimate the minimal number of edges on a path to the sink.  Excess flow can
be \emph{pushed} from a vertex to a neighbor by increasing the flow value on
the connecting edge. Pushes can only happen along \emph{admissible} residual
edges to vertices of lower label.  When none of the edges out of
an active vertex are admissible for a push, the vertex gets \emph{relabeled}
and to a higher label.
It is crucial for practical performance of push-relabel that the labels
estimate the sink distance as accurately as possible.  A simple way to keep
them updated is to regularly run a BFS in the reverse residual graph to set
them to the exact distance. This optimization is called \emph{global
relabeling}.

The more recent \emph{pseudoflow} algorithm due to Hochbaum~\cite{Hochbaum2008} does
not need global relabeling and uses specialized data structures that allow
for pushes along more than one edge.
Implementations of push-relabel algorithms and Hochbaum's algorithm differ
mainly in the order in which they process active vertices. \emph{Highest
label}-based implementations process active vertices in order of decreasing
labels, while FIFO-based implementations select active vertices in queue order.
Goldberg's \emph{hi\_pr} program~\cite{hipr} uses the former technique and is
considered one of the fastest generic maximum flow solvers. It is often used
for comparison purposes in related research.
For push-relabel and Hochbaum's algorithm, it is beneficial to compute merely
a \emph{maximum preflow} that maximizes the cumulative flow on in-edges of the
sink, rather than a complete flow assignment.  In the case where we are looking
only for a minimum cut this is already enough.  In all the other cases,
computing a valid flow assignment for a given maximum preflow can be achieved
using a greedy decomposition algorithm that tends to take much less time than
the computation of a preflow~\cite{Cherkassky1997}.

\subsection{Parallel and Distributed Approaches to the Problem}
\label{sec:relpar}

Parallel algorithms for the maximum flow problem date back to 1982, where Shiloach \emph{et al.}
propose a work-efficient parallel algorithm in the PRAM model, based on blocking flows~\cite{Shiloach1982}.
Most of the more recent work however is based on the push-relabel family of
algorithms. With regard to parallelization, it has the fundamental, distinct
advantage that its primitive operations are inherently local and thus largely
independent.

As far as we know, Anderson and Setubal give the first implementation of
a practical parallel algorithm for the maximum flow problem
\cite{Anderson1995}. In their algorithm, a global queue of active vertices
approximates the FIFO selection order.  A fixed number of threads fetch
vertices from the queue for processing and add newly activated vertices to the
queue using locks for synchronization.
The authors report
speedups over a sequential FIFO push-relabel implementation of up to a factor
of 7 with 16 processors.  The authors also describe a concurrent version of
global relabeling that works in parallel to the asynchronous processing of
active vertices. We will refer to this technique as \emph{concurrent global
relabeling}.
Bader and Sachdeva~\cite{Bader2006} modify the approach by Anderson and Setubal and introduce the first parallel algorithm that approximates the
highest-label vertex selection order used by \emph{hi\_pr}.

Hong~\cite{Hong2008} proposes an asynchronous implementation that completely
removes the need for locking. Instead it makes use of atomic instructions readily
available in modern processors.  Hong and He later present an implementation of
the algorithm that also incorporates concurrent global relabeling
\cite{Hong2011}.  Good speedups over a FIFO-based sequential solver and an
implementation of Anderson and Setubal's algorithm are reported.
There is also a GPU-accelerated implementation of the algorithm~\cite{He2010}.

\emph{Pmaxflow}~\cite{Soner2013} is a parallel, asynchronous FIFO-based
push-relabel implementation. It does not use the concurrent global relabeling
proposed by~\cite{Anderson1995} and instead regularly runs a parallel
breadth-first search on all processors. They report speedups of up to 3 over
\emph{hi\_pr} with 32 threads.

\section{A Synchronous Parallel Implementation of Push-Relabel}

The parallel algorithms mentioned in \autoref{sec:relpar} are exclusively
implemented in an asynchronous manner and differ mainly in the load-balancing
schemes they use and in how they resolve conflicts between adjacent vertices
that are processed concurrently.
We believe the motivation for using asynchronous methods this is that in the tested
benchmark instances, often there is only a handful of active vertices available
for concurrent processing at a given point in time. In this work we try to also
consider larger instances, where there is an obvious need for accelerated processing
and where it might not be possible to solve multiple independent instances concurrently,
due to memory limitations. With a higher number of active vertices per iteration
a synchronous approach becomes more attractive because less work is wasted on distributing the load.

From initial experiments with sequential flow push-relabel algorithms, we
learned the following things: As expected, the average number of active
vertices increases with the size of the graph for a fixed family of inputs.
Also, on almost all of the graphs we tested, a FIFO-based solver outperformed
the highest label-based \emph{hi\_pr} implementation. This is somewhat
surprising as \emph{hi\_pr} is clearly superior on the standard DIMACS
benchmark~\cite{Cherkassky1997}.  These observations led us to an initial
design of a simple synchronous parallel algorithm, inspired by an algorithm
proposed in the original push-relabel article~\cite{Goldberg1988}.  After the
standard initialization, where all edges adjacent to the source are saturated,
it proceeds in a series of iterations, each iteration consisting of the
following steps:

\begin{enumerate}
\item All of the active vertices are processed in parallel. For
each such vertex, its edges are checked sequentially for admissibility.
Possible pushes are performed, but the excess changes are only applied to a copy
of the old excess values. The final values are copied back in step 4.
\item New temporary labels are computed in parallel for
vertices that have been processed in step 1 but are still active.
\item The new labels are applied by iterating again over the set of active vertices
in parallel and setting the distance labels to the values computed in step 2.
\item The excess changes from step 1 are applied by iterating over the
new set of active vertices in parallel.
\end{enumerate}

These steps are repeated until there are no more active vertices with a label
smaller than $n$. The algorithm is deterministic in that it requires the same
amount of work regardless of the number of threads, which is a clear advantage
over other parallel approaches that exhibit a considerable increase in work
when adding more threads~\cite{Hong2011}.  As soon as there are no more active
vertices, we have computed a maximum preflow and can determine a minimum cut
immediately or proceed to reconstruct a maximum flow assignment using
a sequential greedy decomposition.

It is important to note that in step 1 we modify shared memory from
multiple threads concurrently. To ensure correctness, we use atomic fetch-and-add
instructions here to update the excess values of neighbor vertices (or rather,
copies thereof).
Contention on these values is typically low, so overhead caused by cache
coherency mechanisms is not a problem.
To collect the new set of active vertices for the next iteration we use
atomic test-and-set instructions that resolve conflicts when a vertex is activated
simultaneously by multiple neighbors, a situation that occurs only very rarely.
We want to point out that synchronization primitives are kept to a minimum by design,
which to our knowledge constitutes a significant difference to the state-of-the-art.

Instead of running global relabeling concurrently with the rest of the
algorithm as done by~\cite{Anderson1995} and \cite{Hong2011}, we
regularly insert a global relabeling step in between certain iterations. The
work threshold we use to determine when to do this is the same as the one
used by \emph{hi\_pr}.\footnote{Global relabeling is performed approximately
after every $12n + 2m$ edge scans.} The global relabeling is implemented as
a simple parallel reverse breadth-first search from the sink. Atomic compare-and-swap
primitives are used during the BFS to test whether adjacent vertices have already been
discovered.
Apart from global relabeling, we also experimented with other heuristics such as \emph{gap relabeling}, described
in~\cite[Chapter~3]{Cherkassky1997},
but could not achieve speedups by applying them to the parallel case.

\subsection{Improving the Algorithm}
\label{ssec:prsn}

We implemented the above algorithm in
C{}\verb!++! with OpenMP extensions. For common parallel operations like prefix sums and
filter, we used library functions from our \emph{Problem Based Benchmark Suite}
\cite{PBBS} for parallel algorithms.
Even with this very simple implementation, we could measure promising speedups compared
to sequential solvers.  However, we conjectured that the restriction of doing
at most one relabel per vertex in each iteration has some negative consequences:
For one, it hinders the possible parallelism: A low-degree vertex can only
activate so many other vertices before getting relabeled. It would be
preferrable to imitate the sequential algorithms and completely discharge each
active vertex in one iteration by alternating push and relabel operations until
its excess is zero.  Also, the per-vertex work is small. As we
parallelize on a vertex level, we want to maximize the work per vertex to
improve multi-threaded performance in the common case that only few vertices
are active.

To be able to relabel a vertex more than once during one iteration, we need to
allow for non-determinism and develop a scheme to resolve conflicts between
adjacent vertices when both are active in the same iteration. We experimented
with several options here, including the lock-free vertex discharge routine
introduced by Hong and He~\cite{Hong2011}.  Another approach turned out to be more
successful and works without any additional synchronization: In the case where
two adjacent vertices $v$ and $w$ are both active, a deterministic winning
criterion is used to decide which one of the vertices owns the connecting
edges during the current iteration. We say that $v$ wins the competition if
$d(v) < d(w) - 1$ or $d(v) = d(w) + 1$ or $v < w$ (the latter condition is
a tie-breaker in the case where $d(v) = d(w)$). In this case, $v$ is allowed to
push along the edge $(v, w)$ but $w$ is not allowed to push along the edge $(w,
v)$. The discharge of $w$ is thus aborted if $(w, v)$ is the last remaining
admissible edge.  The particular condition is chosen such that one of the
vertices can get relabeled past the other, to ensure progress.  There
is an edge case to consider where two adjacent vertices $v$ and $w$ are
active, $v$ owns the connecting edge but $w$ is still relabeled because the
residual capacity $c_f(w, v)$ is zero. We allow this scenario, but apply
relabels only to a copy of the distance function $d$, called $d'$. The new
admissibility condition for an edge $(x, y) \in E_f$ becomes $d'(x) = d(y)
+ 1$, i.e. the old distance of y is considered. The new labels are applied at
the end of the iteration.

By using this approach, we ensure that for each
sequence of push and relabeling operations in our algorithm during one iteration,
there exists an equivalent sequence of pushes and relabels that is valid with
regard to the original admissibility conditions from~\cite{Goldberg1988}. Thus
the algorithm is correct as per the correctness proof for the push-relabel
algorithm.

The resulting algorithm works similar to the simple algorithm stated above, but
mixes steps 1 and 2, to enable our changes.
We will refer to our own implementation of this algorithm as \emph{prsn} in the
remainder of this document. Pseudocode can be found in the longer version of our paper~\cite{Baumstark2015}.

\section{Evaluation}

\subsection{A Modern Benchmark Suite for Flow Computations}

Traditionally, instance families from the twenty-year-old DIMACS implementation
challenge \cite{Johnson1993} are used to compare the performance of maximum flow
algorithms. Examples of publications that use primarily these graph families are
\cite{Anderson1995,Cherkassky1997,Bader2006,Chandran2009,Goldberg2009,Hong2011}.
We believe that the instance families from the DIMACS benchmark suite do not
accurately represent the flow and cut problems that are typically found today.
Based on different applications where flow computations occur as subproblems,
we compiled a benchmark suite for our experiments that we hope will give us
better insight into which approaches are the most successful in practice.

Saito \emph{et al.} describe how minimum cut techniques can be used for spam
detection on the internet \cite{Saito2007}: They observe that generally spam
sites link to ``good'' (non-spam) sites a lot while the opposite is rarely the
case. Thus the sink partition of a minimum cut between a seed set of good and
spam sites is likely to contain mostly spam sites.  We used their construction
on a graph of pay level domains provided by a research group at the University of
Mannheim with edges of capacity 1 between domains that have at least one
hyperlink.\footnote{\scriptsize\url{http://webdatacommons.org/hyperlinkgraph/}}
A publicly accessible spam
list\footnote{\scriptsize\url{http://www.joewein.de/sw/blacklist.htm}} and a list of major
internet sites\footnote{\scriptsize\url{https://www.quantcast.com/top-sites}} helped us
build good and bad seed sets of size 100 each, resulting in the \emph{pld\_spam} graph.

Very similar constructions can be used for community detection in social
networks~\cite{Flake2000}
It is known that social networks, the Web and document graphs like Wiki\-pedia
share a lot of common characteristics, in particular sparsity and
low diameter. Halim \emph{et al.} include in their article a comprehensive
collection of references that observe these properties for different classes of
graphs~\cite{Halim2011}.  Based on this we believe that \emph{pld\_spam} is
representative of a more general class of applications that involve community
detection in such graphs.

Graph partitioning software such as \emph{KaHIP} due to Sanders and Schulz
commonly use flow techniques internally~\cite{Sanders2011}.
The KaHIP website\footnote{\scriptsize\url{http://algo2.iti.kit.edu/documents/kahip/}}
provides an archive of flow instances for research purposes which
we used as part of our test suite. We included multiple instances from this
suite, because the structure of the flow graphs is very close to the structure
of the input graphs and those cover a wide range of practical applications.

The input graphs for KaHIP are taken from the 10th DIMACS graph partitioning
implementation challenge~\cite{Dimacs2012}: \emph{delaunay} is a family of
graphs representing the Delaunay triangulations of randomly generated sets of
points in the plane.  \emph{rgg} is a family of random geometric graphs
generated from a set of random points in the unit square.  Points are connected
via an edge if their distance is smaller than $0.55\cdot\frac{\ln n}{n}$.
\emph{europe.osm} is the largest amongst a set of street map graphs.
\emph{nlpkkt240} is the graph representation of a large sparse matrix arising
in non-linear optimization.
For the cases where graphs of different sizes are available (\emph{delaunay} and
\emph{rgg}), we included the largest instance whose internal representation fits
into the main memory of our test machine.

As a third application, in computer vision a lot of different problems reduce
to minimum cut: For reference, Fishbain and Hochbaum \cite[Section
3.2]{Fishbain2013} describe various examples of applications.
We included \emph{BL06-camel-lrg}, an instance of multi-view reconstruction from
the vision benchmark suite of the University of Western
Ontario.\footnote{\scriptsize\url{http://vision.csd.uwo.ca/data/maxflow/}}

For completeness, we also included instances of two of the harder graph families
from the DIMACS maximum flow challenge, \emph{rmf\_wide\_4} and \emph{rlg\_wide\_16}, which
are described for example by~\cite{Cherkassky1997}.
\autoref{tab:bench_sizes} shows the complete list of graphs we used in our
benchmarks, together with their respective vertex and edge counts, as well as
the maximum edge capacities.

\vspace{-1em}
\begin{table}
\centering
\caption{Properties of our benchmark graph instances. The maximum edge capacity
is excluding source or sink adjacent edges.}
\label{tab:bench_sizes}
\begin{tabular}{| l | r | r | r |}
  \hline
  \textbf{graph name} & \textbf{num. vertices} & \textbf{num. edges} & \textbf{max. edge capacity}\\ \hline
  rmf\_wide\_4 & \num{1048576} & \num{5160960} & 10000 \\
  rlg\_wide\_16 & \num{4194306} & \num{12517376} & 30000 \\
  delaunay\_28 & \num{161061274} & \num{966286764} & 1\\
  rgg\_27 & \num{80530639} & \num{1431907505} & 1\\
  europe.osm & \num{15273606} & \num{32521077} & 1\\
  nlpkkt240 & \num{8398082} & \num{222847493} & 1\\
  pld\_spam & \num{42889802} & \num{623056513} & 1\\
  BL06-camel-lrg & \num{18900002} & \num{93749846} & 16000\\
  \hline
\end{tabular}
\end{table}
\vspace{-2em}
\subsection{Comparison and Testing Methodology}

Our aim was to compare the practical efficiency of our algorithm to the
sequential and parallel state-of-the-art on a common Intel architecture.
For comparison with sequential implementations, we selected the publicly available
\emph{f\_prf}\footnote{\scriptsize\url{http://www.avglab.com/soft.html}},
\emph{hi\_pr} and
\emph{hpf}\footnote{\scriptsize\url{http://riot.ieor.berkeley.edu/Applications/Pseudoflow/maxflow.html}}
programs, implementing FIFO and highest label-based push-relabel and
Hochbaum's pseudoflow algorithm, respectively. For \emph{hi\_pr}, we did not find
a canonical URL for the most recent 3.7 version of the code and instead used the
copy embedded in a different
project.\footnote{\scriptsize\url{https://code.google.com/p/pmaxflow/source/browse/trunk/goldberg/hipr}}
Our results show that \emph{hpf} is the best sequential solver
for our benchmark suite, only outperformed by \emph{f\_prf} on the \emph{pld\_spam}
graph.
The most recent parallel algorithm is the asynchronous lock-free algorithm by
Hong and He~\cite{Hong2011}. Since it has no public implementation, we
implemented their algorithm based on the pseudocode description.  We will refer to
it as \emph{hong\_he} in the remainder of this document. Our own implementation
of the algorithm described in \autoref{ssec:prsn} is called \emph{prsn}.
We also experimented with a parallel push-relabel implementation
that is part of the Galois
project.\footnote{\scriptsize\url{http://iss.ices.utexas.edu/?p=projects/galois/benchmarks/preflow_push}}
Although their code is competitive on certain small inputs,
it did not complete within a reasonable amount of time on larger instances.

To eliminate differences due to graph representation and initialization
overhead, we modified the algorithms to use the same internal graph representation,
namely adjacency arrays with each edge also storing the residual capacity of its
reverse edge, as described in~\cite{Bader2006}.
For all five algorithms we only measured the time to compute the
maximum preflow, not including the data structure initialization time.  The
reconstruction of the complete flow function from there
is the same in every case and takes only a negligible fraction (less than
3 percent) of the total sequential computation time for all of our input
graphs. We measured each combination of algorithm and input at least five
times.

We carried out the experiments on a NUMA Intel Nehalem machine. It hosts four
Xeon E7-8870 sockets clocked at 2.4 GHz per core, making for a total of 40 physical
and 80 logical cores.  Every socket has 64 GiB of RAM associated with it, making for
a total of 256 GiB.

\vspace{-1em}
\subsection{Results}
\label{subsec:results}

\vspace{-1.5em}
\begin{figure}
\begin{minipage}[t]{0.47\textwidth}
\hspace*{-1em}
\includegraphics[width=1.1\columnwidth]{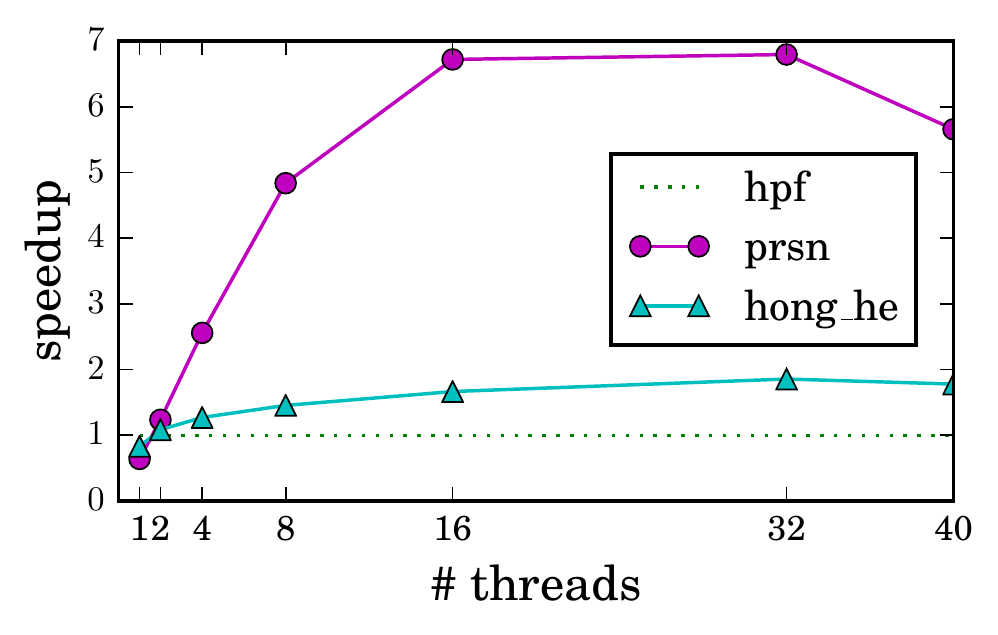}
\vspace{-2.5em}
\caption{\label{fig:speedup_rgg_27}Speedup for \emph{rgg\_27} compared to the
best single-threaded timing.}
\end{minipage}\hfill
\begin{minipage}[t]{0.47\textwidth}
\hspace*{-1em}
\includegraphics[width=1.1\columnwidth]{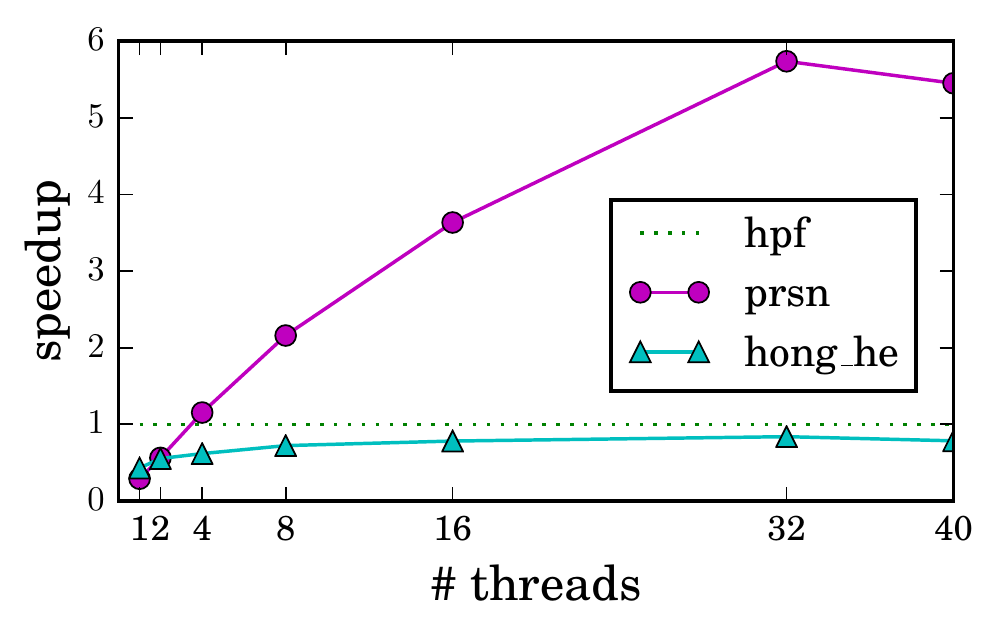}
\vspace{-2.5em}
\caption{\label{fig:speedup_nlpkkt240}Speedup for \emph{nlpkkt240} compared to the
best single-threaded timing.}
\end{minipage}
\end{figure}
\vspace{-1em}

\begin{figure}
\begin{minipage}[t]{0.47\textwidth}
\hspace*{-1em}
\includegraphics[width=1.1\columnwidth]{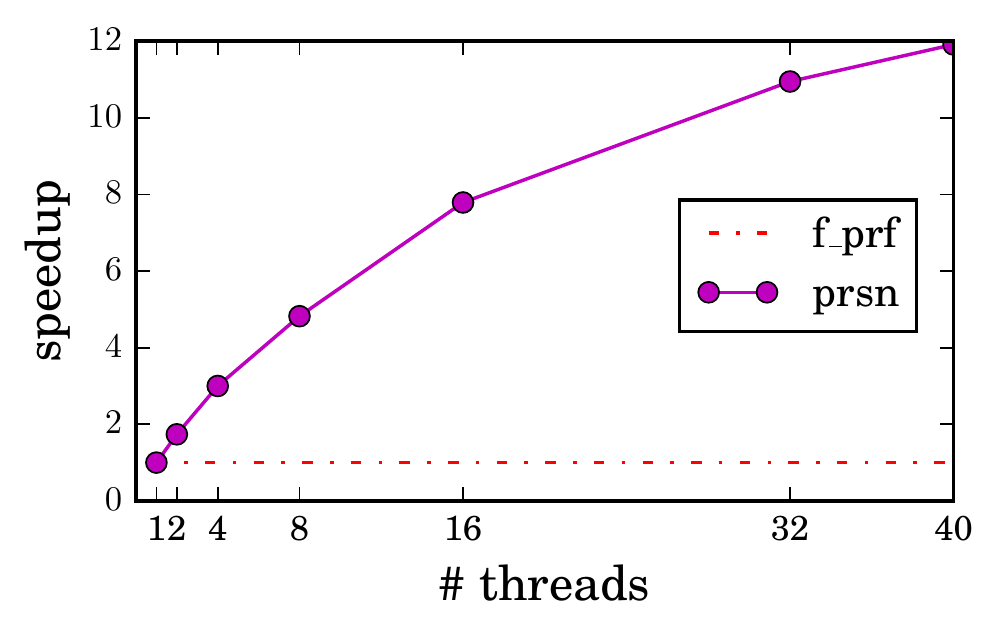}
\vspace{-2.5em}
\caption{\label{fig:speedup_pld_spam}Speedup for \emph{pld\_spam} compared to the
best  timing. \emph{hong\_he} did not
finish in our experiments.} \end{minipage}\hfill
\begin{minipage}[t]{0.47\textwidth}
\hspace*{-1em}
\includegraphics[width=1.1\columnwidth]{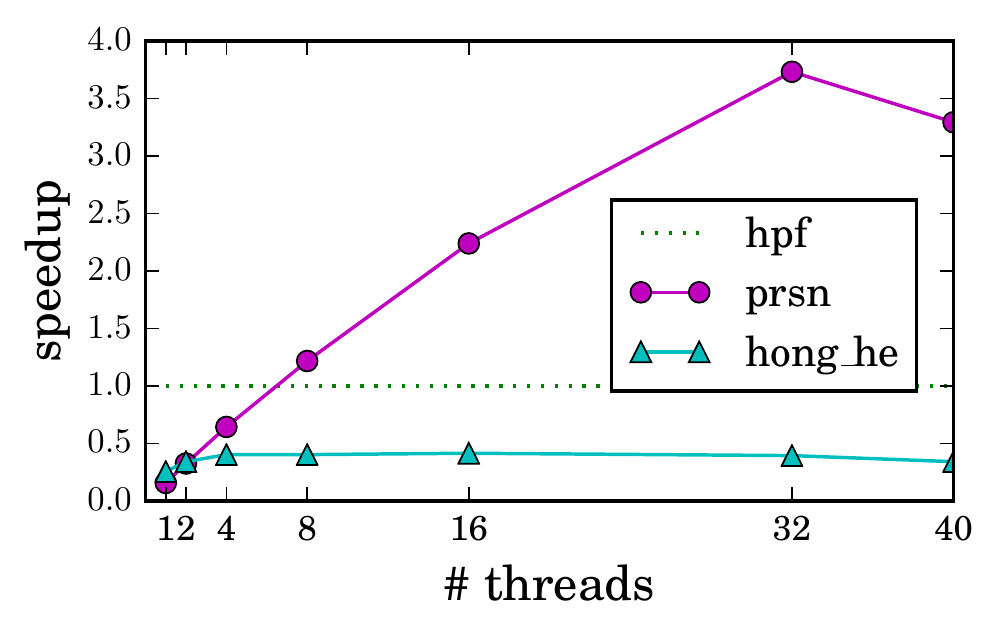}
\vspace{-2.5em}
\caption{\label{fig:speedup_BL06-camel-lrg}Speedup for \emph{BL06-camel-lrg}
compared to the best single-threaded timing.}
\end{minipage}
\end{figure}

The longer version of our paper contains comprehensive tables of the absolute timings
we collected in all our experiments~\cite{Baumstark2015}.
\emph{rgg\_27}, \emph{delaunay\_28} and \emph{nlpkkt240} are examples of graphs
where an effective parallel solution is possible: \autoref{fig:speedup_rgg_27}
shows that both \emph{hong\_he} and \emph{prsn} outperform \emph{hpf} in the
case of \emph{rgg\_27}; furthermore \emph{prsn} is three times faster than
\emph{hong\_he} with 32 threads. The speedup plot for \emph{delaunay\_28} looks almost
identical to the one for \emph{rgg\_27}. In the case
of \emph{nlpkkt240}, we can tell from \autoref{fig:speedup_nlpkkt240} that
\emph{prsn} outperforms \emph{hpf} with four threads and achieves
a speedup of 5.7 over \emph{hpf} with 32 threads. \emph{hong\_he} does not
achieve any absolute speedup even with 40 threads.
\emph{prsn} does remarkably well on our spam detection instance
\emph{pld\_spam}: Even with one thread, our implementation outperforms \emph{hpf} and
\emph{hi\_pr} and is on par with \emph{f\_prf}. \autoref{fig:speedup_pld_spam} shows that with 40 threads, an
absolute speedup of 12 is achieved over the best sequential run. We
noticed here that the algorithm spends most of the time performing a small
number of iterations on a very large number of active vertices, which is very
advantageous for parallelization. Note that \emph{hong\_he} did not finish on
the \emph{pld\_spam} benchmark after multiple hours of run time. We conjecture
that this is because the algorithm is simply very inefficient for this
particular instance and were able to confirm that this is the case by
reimplementing the same vertex discharge in the sequential \emph{f\_prf}
program. Before each push, it scans all the edges of a vertex to find the
neighbor with the lowest label. This is necessary in \emph{hong\_he} because
the algorithm does not maintain the label invariant $d(x) \leq d(y) + 1$ for
all $(x, y) \in E_f$. The modified \emph{f\_prf} also did not finish solving
the benchmark instance within a reasonable time frame.

\vspace{-1.5em}
\vspace{-1.5em}

\emph{BL06-camel-lrg} is a benchmark from computer vision.
\autoref{fig:speedup_BL06-camel-lrg} shows that \emph{prsn} is able to outperform
\emph{hpf} with 8 threads and achieves a speedup of almost four with 32
threads.  \emph{hpf} has in turn been shown to perform almost as well as the
specialized BK algorithm on this benchmark \cite{Fishbain2013}.

\begin{figure}
\begin{minipage}[t]{0.47\textwidth}
\hspace*{-1em}
\includegraphics[width=1.1\columnwidth]{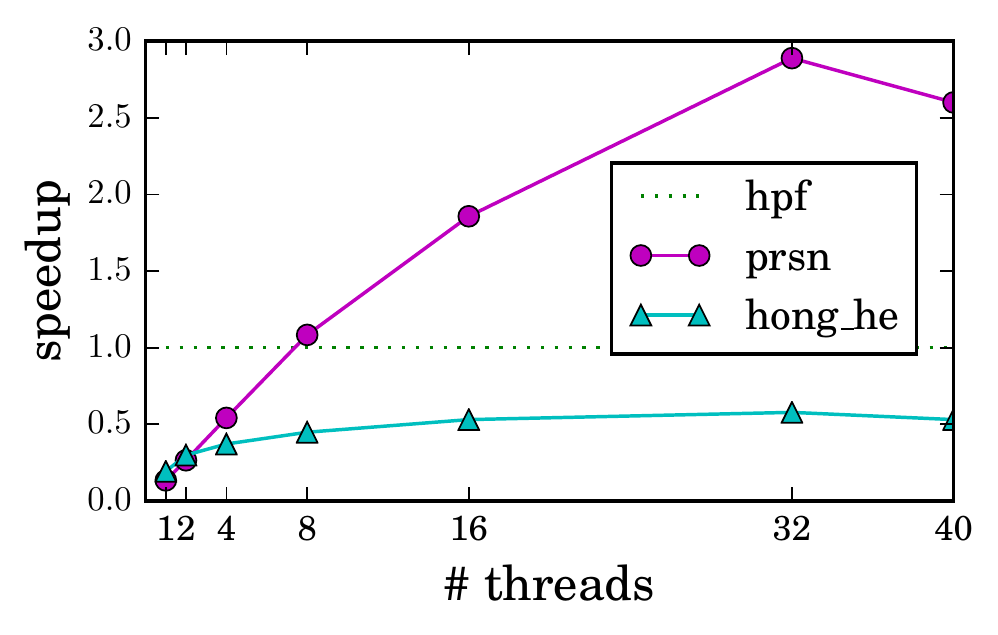}
\vspace{-2.5em}
\caption{\label{fig:speedup_washington_rlg_wide_16}Speedup for
\emph{rlg\_wide\_16} compared to the best sequential timing.}
\end{minipage}\hfill
\begin{minipage}[t]{0.47\textwidth}
\hspace*{-1em}
\includegraphics[width=1.1\columnwidth]{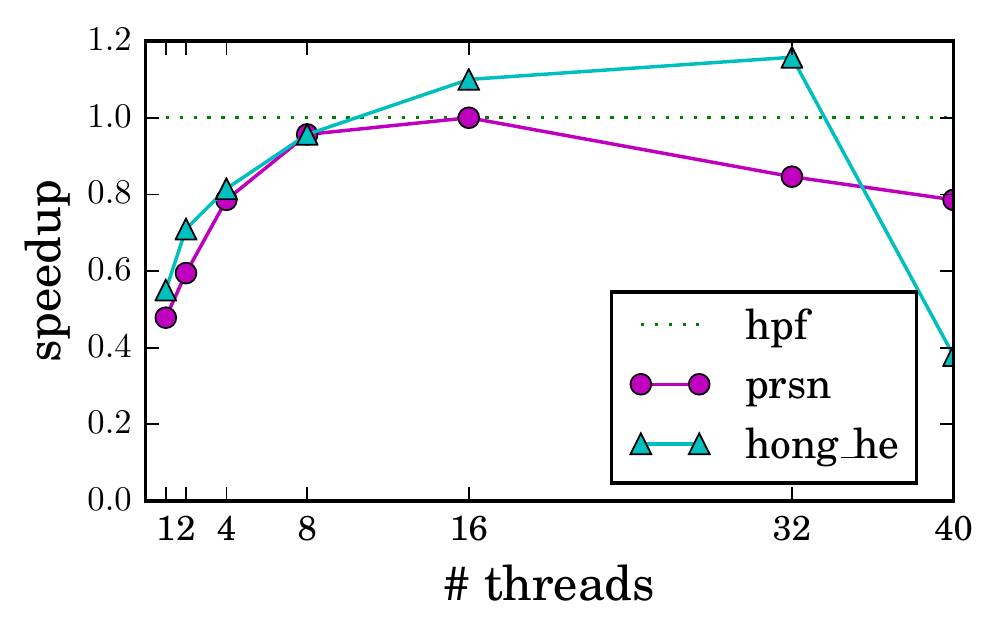}
\vspace{-2.5em}
\caption{\label{fig:speedup_europe_osm}Speedup for \emph{europe.osm} compared to
the best single-threaded timing.}
\end{minipage}
\end{figure}

As we can tell from \autoref{fig:speedup_washington_rlg_wide_16}, in the case of
\emph{rlg\_wide\_16}, \emph{prsn} requires eight threads to outperform
\emph{hpf} and achieves an absolute speedup of about three with 32 threads.
\emph{europe.osm} appears to be a hard instance for the parallel algorithms, as
shown in~\autoref{fig:speedup_europe_osm}: Only \emph{hong\_he} achieves a small
speedup with 32 threads.
Both parallel algorithms fail to outperform the best sequential algorithm in the case
of the \emph{rmf\_wide\_4} graph.
In all cases, making use of hyper-threading by running the algorithms with 80 threads
did not yield any performance improvements, which we attribute to the fact
that the algorithm is mostly memory bandwidth-bound.

Overall, different graph types lead to different behaviour of the tested algorithms.
We have shown that especially for large, sparse graphs of low diameter, our
algorithm can provide significant speedups over existing sequential and parallel
maximum flow solvers.
\section{Conclusion}

In this paper, we presented a new parallel maximum flow implementation and
compared it with existing state-of-the-art sequential and parallel
implementations on a variety of graphs. Our implementation uses coarse-grained
synchronization to avoid the overhead of fine-grained locking and
hardware-level synchronization used by other parallel implementations. We showed
experimentally that our implementation outperforms the fastest existing parallel
implementation and achieves good speedup over existing sequential
implementations on different graphs. Therefore, we believe that our algorithm can
considerably accelerate many flow and cut computations that arise in practice.
To evaluate the performance of our algorithm, we identified a new set of benchmark
graphs representing maximum flow problems occuring in practical applications. We
believe this contribution will help in evaluating maximum
flow algorithms in the future.

\vspace{-1em}
\subsubsection*{Acknowledgments.}

We want to thank Professor Dr. Peter Sanders from Karlsruhe Institute of Technology
for contributing initial ideas and Dr. Christian Schulz from KIT for preparing flow
instances for us.

This work is partially supported by the National Science Foundation under
grant CCF-1314590, and by the Intel Labs Academic Research Office for the
Parallel Algorithms for Non-Numeric Computing Program.

\bibliography{references}

\clearpage
\section*{Appendix: Pseudocode and Timings}

\vspace{-1em}
\renewcommand\mylstcaption{Pseudocode implementation of \emph{prsn}.}
\begin{lstlisting}[caption=\mylstcaption, label=code:prsn]
procedure PRSyncNondet()
    parallel foreach %*$v \in V$*)
        d(v) := 0
        e(v) := 0
        v.addedExcess := 0
        v.isDiscovered := 0
    d(s) := n
    parallel foreach %*$(v, w) \in E$*)
        f(v,w) := f(w,v) := 0
    // initially saturate all source-adjacent edges
    parallel foreach %*$(s, v) \in E$*)
        f(s,v) := c(s,v)
        f(v,s) := -c(s,v)
        e(v) := c(s,v)
    workSinceLastGR := %*$\infty$*)
    while true:
        // from hi_pr: freq = 0.5, %*$\alpha = 6$*)
        if freq %*$\cdot$*) workSinceLastGR > %*$\alpha \cdot n + m$*):
            workSinceLastGR := 0
            GlobalRelabel()  // see %*\autoref{code:GlobalRelabel}*)
            // parallel array comprehension using map/filter
            workingSet = [ v | v %*$\leftarrow$*) workingSet, d(v) < n ]

        if workingSet = %*$\emptyset$*) break

        parallel foreach %*$v \in$*) workingSet
            v.discoveredVertices := []
            d%*'*)(v) := d(v)
            e := e(v) // local copy
            v.work := 0
            while e > 0
                newLabel := n
                skipped := 0
                parallel foreach residual edge %*$(v, w) \in E_f$*)
                    if e = 0   // vertex is already discharged completely
                        break
                    admissible := (d%*'*)(v) = d(w) + 1)
                    // is the edge shared between two active vertices?
                    if e(w)
                        win := d(v) = d(w) + 1
                                    or d(v) < d(w) - 1
                                    or (d(v) = d(w) and v < w)
                        if admissible and not win
                            skipped := 1
                            continue   // skip to next residual edge
                    if admissible and %*$c_f(v, w) > 0$*) // edge is admissible
                        %*$\Delta$*) := %*$\min(c_f(v, w), e(v))$*)
                        // the following three updates do not need to be atomic
                        f(v,w) += %*$\Delta$*)
                        f(w,v) -= %*$\Delta$*)
                        e -= %*$\Delta$*)
                        // atomic fetch-and-add
                        w.addedExcess += %*$\Delta$*)
                        if w %*$\neq$*) t and TestAndSet(w.isDiscovered)
                            v.discoveredVertices.pushBack(w)
                    if %*$c_f(v, w) > 0$*) and %*$d(w) \geq d'(v)$*)
                        newLabel := %*$\min$*)(newLabel, d(w) + 1)
                if e = 0 or skipped
                    break
                d%*'*)(v) := newLabel // relabel
                v.work += v.outDegree + %*$\beta$*)  // from hi_pr: %*$\beta = 12$*)
                if d%*'*)(v) = n
                    break
            v.addedExcess := e - e(v)
            if e%*'*)(v) and TestAndSet(v.isDiscovered)
                v.discoveredVertices.pushBack(v)

        parallel foreach %*$v \in$*) workingSet
            d(v) := d%*'*)(v)
            e(v) += v.addedExcess
            v.addedExcess := 0
            v.isDiscovered := 0

        workSinceLastGR += Sum([ v.work | v %*$\leftarrow$*) workingSet ])
        workingSet := Concat([ v.discoveredVertices | v %*$\leftarrow$*) workingSet, d(v) < n ])

        parallel foreach %*$v \in$*) workingSet
            e(v) += v.addedExcess
            v.addedExcess := 0
            v.isDiscovered := 0
\end{lstlisting}

\vspace{-1em}
\renewcommand\mylstcaption{Pseudocode implementation of parallel global relabeling.}
\begin{lstlisting}[caption=\mylstcaption, label=code:GlobalRelabel]
procedure GlobalRelabel()
    parallel foreach %*$v \in V$*)
        d(v) := n
    d(t) := 0
    Q := [t]
    while Q %*$\neq \emptyset$*)
        parallel foreach %*$v \in Q$*)
            v.discoveredVertices := []
            for each edge %*$(v, w) \in E_f$*) with %*$w \neq s$*) and %*$c_f(v, w) > 0$*)
                // this branch must be implemented atomically using %*\\*) compare-and-swap
                if %*$w \neq t$*) and %*$d(w) = n$*)
                    d(w) := d(v) + 1
                    v.discoveredVertices.pushBack(w)
        // concatenation implemented using parallel prefix sums
        Q := Concat([ v.discoveredVertices | v %*$\leftarrow$*) Q ])
\end{lstlisting}

\vspace{-1em}
\begin{table}[]
\centering
\caption{Sequential benchmark results. The numbers represent the time in
seconds to find a maximum preflow (excluding initialization of the graph
data structure). Timings are averaged over at least 5 runs. For each row, the
best timing is marked in bold. The cell marked with ``DNF'' represents
an experiment that did not finish. Please refer to~\autoref{subsec:results} for
our analysis of why these runs failed.}
\begin{tabular}{| l | r | r | r | r | r |}
  \hline
  \textbf{graph} & \textbf{hi\_pr} & \textbf{hpf} & \textbf{f\_prf} & \textbf{prsn} & \textbf{hong\_he}
                 \\ \hline
genrmf\_wide\_4 & 41 & \textbf{11} & 81 & 260 & 228 \\
washington\_rlg\_wide\_16 & 88 & \textbf{26} & 118 & 194 & 135 \\
delaunay\_28 & 21564 & \textbf{2905} & 8124 & 4665 & 4112 \\
rgg\_27 & 13082 & \textbf{2433} & 6937 & 3807 & 2929 \\
europe\_osm & 359 & \textbf{22} & 76 & 46 & 40 \\
nlpkkt240 & 521 & \textbf{218} & 711 & 752 & 508 \\
pld\_spam & 443 & 907 & \textbf{405} & \textbf{405} & DNF \\
BL06-camel-lrg & 259 & \textbf{56} & 220 & 358 & 219 \\
  \hline
\end{tabular}
\label{tab:results_seq}
\end{table}

\vspace{-1em}
\begin{table}[]
\centering
\caption{Parallel benchmark results for \emph{prsn}. The numbers represent the
time in seconds to find a maximum preflow (excluding initialization of the
graph data structure). Timings are averaged over at least 5 runs.}
\begin{tabular}{| l | r | r | r | r | r | r | r |}
  \hline
  \multirow{2}{*}{\textbf{graph}}
  &\multicolumn{7}{c|}{\textbf{number of threads}}\\
  \cline{2-8}
  & \textbf{1} & \textbf{2} & \textbf{4} & \textbf{8} & \textbf{16} & \textbf{32} & \textbf{40} \\ \hline
  genrmf\_wide\_4 & 260 & 202 & 139 & 102 & 93 & 109 & 119 \\
washington\_rlg\_wide\_16 & 194 & 98 & 48 & 24 & 14 & 9 & 10 \\
delaunay\_28 & 4665 & 2151 & 1180 & 619 & 560 & 493 & 491 \\
rgg\_27 & 3807 & 1970 & 951 & 503 & 362 & 358 & 430 \\
europe\_osm & 46 & 37 & 28 & 23 & 22 & 26 & 28 \\
nlpkkt240 & 752 & 388 & 189 & 101 & 60 & 38 & 40 \\
pld\_spam & 405 & 233 & 135 & 84 & 52 & 37 & 34 \\
BL06-camel-lrg & 358 & 172 & 87 & 46 & 25 & 15 & 17 \\
  \hline
\end{tabular}
\label{tab:results_prsn}
\end{table}

\vspace{-1em}
\begin{table}[]
\centering
\caption{Parallel benchmark results for \emph{hong\_he}. The numbers represent
the time in seconds to find a maximum preflow (excluding initialization of the
graph data structure). Timings are averaged over at least 5 runs. The cells
marked with ``DNF'' represent experiments that did not finish. Please refer
to~\autoref{subsec:results} for our analysis of why these runs failed.}
\begin{tabular}{| l | r | r | r | r | r | r | r |}
  \hline
  \multirow{2}{*}{\textbf{graph}}
  &\multicolumn{7}{c|}{\textbf{number of threads}}\\
  \cline{2-8}
  & \textbf{1} & \textbf{2} & \textbf{4} & \textbf{8} & \textbf{16} & \textbf{32} & \textbf{40} \\ \hline
genrmf\_wide\_4 & 228 & 121 & 97 & 86 & 50 & 41 & 61 \\
washington\_rlg\_wide\_16 & 135 & 87 & 70 & 58 & 49 & 45 & 49 \\
delaunay\_28 & 4112 & 2414 & 2383 & 2313 & 1888 & 1590 & 1553 \\
rgg\_27 & 2929 & 2245 & 1915 & 1673 & 1461 & 1311 & 1368 \\
europe\_osm & 40 & 31 & 27 & 23 & 20 & 19 & 58 \\
nlpkkt240 & 508 & 394 & 353 & 302 & 279 & 260 & 278 \\
pld\_spam & DNF & DNF & DNF & DNF & DNF & DNF & DNF \\
BL06-camel-lrg & 219 & 164 & 139 & 139 & 135 & 142 & 164 \\
  \hline
\end{tabular}
\label{tab:results_hong_he}
\end{table}

\vspace{-1em}
\begin{figure}
\begin{minipage}[t]{0.47\textwidth}
\hspace*{-1em}
\includegraphics[width=1.1\columnwidth]{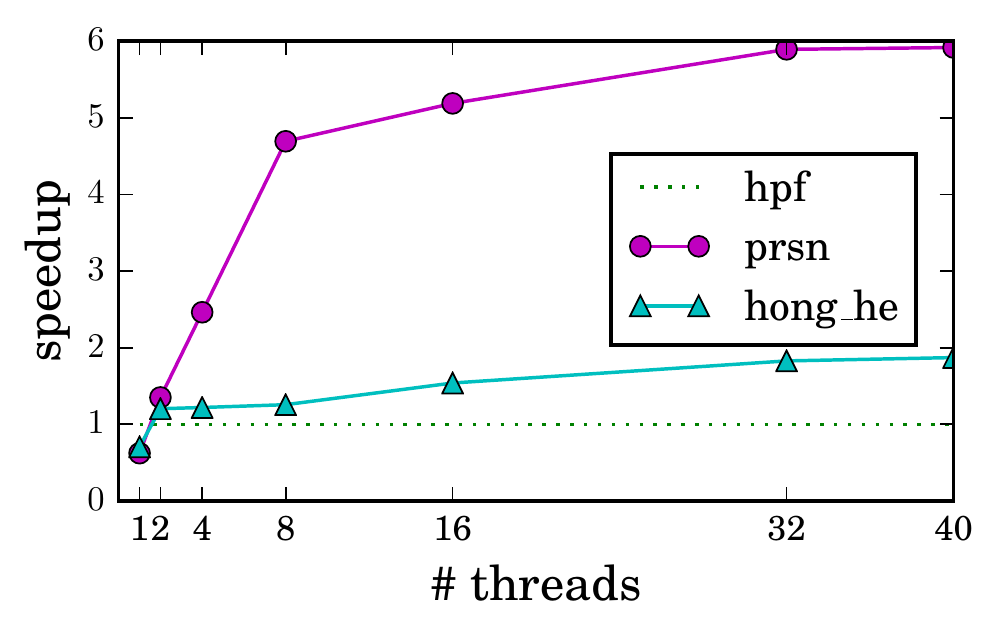}
\vspace{-2.5em}
\caption{\label{fig:speedup_delaunay_28}Speedup for \emph{delaunay\_28} compared to the best
single-threaded timing.}
\end{minipage}\hfill
\begin{minipage}[t]{0.47\textwidth}
\hspace*{-1em}
\includegraphics[width=1.1\columnwidth]{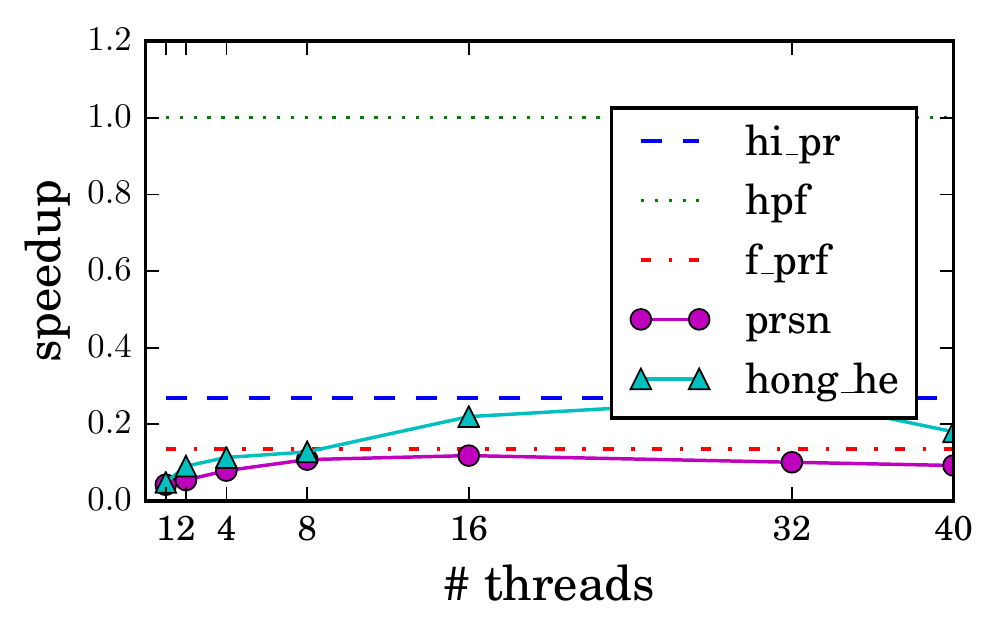}
\vspace{-2.5em}
\caption{\label{fig:speedup_genrmf_wide_4}Speedup for \emph{genrmf\_wide\_4}
compared to the best single-threaded timing.}
\end{minipage}
\end{figure}

\end{document}